\documentclass[]{aa}  

\usepackage{graphicx}
\usepackage{txfonts}

\newcommand{\INTEGRAL}{{\it INTEGRAL}}

\newcommand{\refisgri}{\citep[\INTEGRAL\/ Soft Gamma-Ray Imager,][]{lebrun03}}

\newcommand{\refspi}{(SPectrometer on INTEGRAL)}

\begin{document} 

\title{GRB\,190919B: Rapid optical rise explained as a flaring activity}
\author{
    Martin Jelínek\inst{1} \fnmsep\thanks{\email{mates@asu.cas.cz}}
    \and
    Martin Topinka\inst{2,3}
    \and
    Sergey Karpov\inst{4}
    \and
    Alžběta Maleňáková\inst{1,5}
    \and
    Y.-D. Hu\inst{6} 
    \and
    Michela Rigoselli\inst{2}
    \and 
    Jan \v{S}trobl\inst{1}
    \and
    Jan Ebr\inst{4}
    \and
    Ronan Cunniffe\inst{4}
    \and
    Christina Thoene\inst{1}
    \and
    Martin Ma\v{s}ek\inst{4}
    \and
    Petr Jane\v{c}ek\inst{4}
    \and 
    Emilio Fernandez-García\inst{6}
    \and
    David Hiriart\inst{7}
    \and 
    William H. Lee\inst{8}
    \and
    Stanislav Vítek\inst{9}
    \and
    René Hudec\inst{1,9}
    \and 
    Petr Trávní\v{c}ek\inst{4}
    \and 
    Alberto J. Castro-Tirado\inst{6}
    \and
    Michael Prouza\inst{4}
    }

\institute{  
    Astronomical Institute (ASU CAS), Ondřejov, Czech Republic \email{mates@asu.cas.cz} 
    \and
    INAF – Istituto di Astrofisica Spaziale e Fisica Cosmica, 
    Milano, Italy
    \and
     Department of Theoretical Physics and Astrophysics, Masaryk University,
     Brno, Czech Republic       
     \and
    Institute of Physics of Czech Academy of Sciences, 
    Prague, Czech Republic
    \and
    Astronomical Institute, MFF UK, Prague, Czech Republic 
    \and
     Instituto de Astrofísica de Andalucía (IAA-CSIC), 
     Granada, Spain
     \and
     Instituto de Astronom\'ia de la UNAM, 
     Ensenada, Baja Cfa., M\'exico.
     \and
     Circuito Exterior s/n, Ciudad Universitaria, Delg. Coyoac\'an, M\'exico D.F.,
    M\'exico
     \and
     Faculty of Electrical Engneering (FEL-CVUT), 
     Prague, Czech Republic
}

\titlerunning{GRB\,190919B: Rapid optical rise explained as a flaring activity}
\authorrunning{Jel\'{\i}nek et al.}

\date{Received December 27, 2021 ; accepted March 8, 2022}
   
\abstract{
Following the detection of a long GRB\,190919B by INTEGRAL (INTErnational Gamma-Ray Astrophysics Laboratory), we obtained an optical photometric sequence of its optical counterpart. The light curve of the optical emission exhibits an unusually steep rise $\sim 100$\,s after the initial trigger. This behaviour is not expected from a
'canonical' GRB optical afterglow. As an explanation, we propose a scenario
consisting of two superimposed flares: an optical flare originating from the
inner engine activity followed by the hydrodynamic peak of an external shock.
The inner-engine nature of the first pulse is supported by a marginal detection
of flux in hard X-rays. The second pulse
eventually concludes in a slow constant decay, which, as we show, follows the
closure relations for a slow cooling plasma expanding into the constant
interstellar medium and can be seen as an optical afterglow \emph{sensu
stricto}.}
   

   \maketitle

\section{Introduction}

Gamma-ray bursts (GRBs) are undoubtedly the brightest single events occurring at
cosmological distances. Their isotropic luminosity may reach up to
$10^{54}$\,erg\,s$^{-1}$. A broadly accepted scheme explains the prompt
gamma-ray emission as produced by internal dissipation processes within the
relativistic ejecta \citep{Piran2004, Meszaros2006, Zhang2007, KumarZhang2015}.
The long-lived broadband afterglow emission is then usually described in terms
of an interaction of the ultra-relativistic ejecta with the ambient medium
\citep{MeszarosRees1997,Sari1998}.

The growing ability of rapid follow-up by both X-ray and optical and near-infrared instruments (large number of both space- and ground-based telescopes)
permitted us to see the transition between these two modes. A world of
interesting features was found in both optical and X-rays as well as in the
relation between the two. In contrast to a rather humdrum afterglow behaviour
at later times, in the early light curves one may see the onset of the emission and
breaks to steeper or shallower decay and flares \citep{Zaninoni2013,
Swenson2013, Yi2017}.

In this paper, we present observations of the onset of the afterglow of
GRB\,190919B and propose an explanation for its far too steeply rising
afterglow emission. This is something the relativistic fireball model of an afterglow
can not explain.
\begin{figure}[b!]
    \centering
    \includegraphics[width=\hsize]{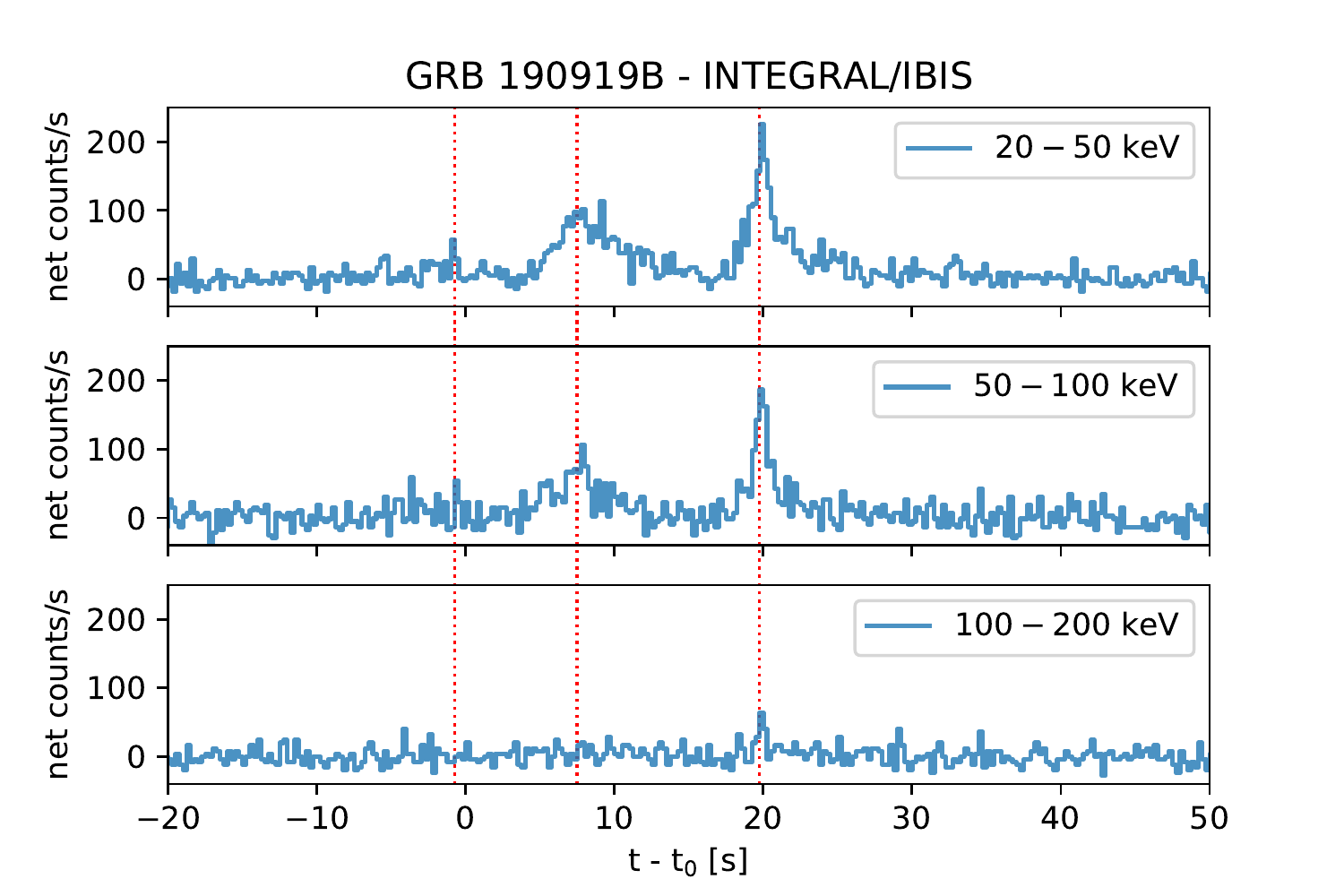}
    \caption{INTEGRAL/ISGRI $20 - 50$~keV and $50-150$~keV light curves
of GRB 190919B. The vertical lines depict the times of individual flare maxima
in the high energy range. The hard energy excess is detected during the second
main flare at $(t-t_0) \sim 20$~s in the 100-200~keV energy
range.\label{fig:int_2lc}}
\end{figure}

\section{Observations}

The long GRB,\,190919B was detected by \INTEGRAL/IBAS \citep[INTErnational
Gamma-Ray Astrophysics Laboratory, INTEGRAL Burst Alert System][]{winkler03,
kuulkers21,mereghetti03} on September 19, 2019 at 23:46:40 UT in the southern
constellation of The Microscope \citep{GCN25788}. 

The event localisation was available within 34.6\,s to 1.5' accuracy. This
precision permitted rapid ground-based follow-up with a number of telescopes,
and soon the optical counterpart was discovered at 
\begin{flushright} 
$\alpha$ = 20:47:30.615 \quad $ \delta $ = -44:41:43.03 \qquad (J2000)
\end{flushright}
\noindent\citep{GCN25789}, and the redshift was determined from several absorption
lines in the afterglow spectrum as $z=3.225$ \citep{GCN25792}.

As found at the Burst analyser
web page\footnote{https://www.swift.ac.uk/bu\-rst\_a\-na\-ly\-ser/}
\citep{Evans2007,Evans2009,Evans2010}, GRB\,190919B was observed in
X-rays for 3\,ks by {\it Swift}\/-XRT (X-Ray Telescope) $\sim$30\,ks after the trigger. We
adopted the results of the spectral fit for this observation, and for reference
copy them in Table\,\ref{tab:xrt}. A further observation was performed
$\sim$128\,ks after trigger, but the signal was too faint to centre a centroid
at the object, and the pipeline marks this observation as unreliable. Despite this, the Burst analyser mentions an X-ray decay rate of $\alpha_\mathrm{X} =
1.30{^{+0.5}_{-0.4}}$ based on the two mentioned observing epochs.


\begin{table}[t!]
\begin{center}
\begin{tabular}{lc}
\hline
N$_{\mathrm{H}}$ (Galactic)     & $2.75 \times 10^{20}$ cm$^{-2}$ \\
N$_{\mathrm{H}}$ (intrinsic)  & $2.3 {^{+4.4}_{-2.3}} \times 10^{22}$ cm$^{-2}$ \\
z of absorber & 0 \\
Photon index &  $2.1^{+0.6}_{-0.5}$ \\
Flux (0.3-10 keV) 
(Obs.) &        $7.6{^{+3.1}_{-2.1}} \times 10^{-13}$ erg cm$^{-2}$ s$^{-1}$ \\
Flux (0.3-10 keV) 
(Unabs.)&       $9.6{^{+3.3}_{-2.3}} \times  10^{-13}$ erg cm$^{-2}$ s$^{-1}$ \\
Counts to flux (obs) &  $3.17 \times 10^{-11}$ erg cm$^{-2}$ ct$^{-1}$ \\
Counts to flux (unabs) & $4.00 \times 10^{-11}$ erg cm$^{-2}$ ct$^{-1}$ \\
W-stat (dof) &  53.58 (47) \\
Spectrum exposure       & 3.0\,ks \\
Mean photon arrival & T0+31417 s \\
\hline
\end{tabular}
\end{center}
\caption{Parameters of an X-ray spectral fit of the 3\,ks observation by {\it
Swift}\/-XRT. Adopted without changes from the Burst analyser
\citep{Evans2010} {\tt
https://www.swift.ac.uk/xrt\_live\_cat/00020948/}\label{tab:xrt}}
\end{table}

%
%


\subsection{INTEGRAL}

Of the instruments aboard the spacecraft, INTEGRAL/ISGRI {\refisgri} was the
primary detector to provide data for our analysis, and SPI {\refspi} provided
some extra signal to be included in the processing, despite reduced performance
due to annealing. The GRB was 8.9$^\circ$ off the main axis of the spacecraft,
away from the fields of view of Joint European X-Ray Monitor (JEM-X) and the Optical Monitoring Camera (OMC). The high-energy photon detector PICsIT and the anti-coincidence
shield did not provide any useful data either. 

We processed the data from the INTEGRAL archive with the standard software (Offline Scientific Analysis, OSA
version 11). The GRB consisted of a dim first pulse followed by two bright
emission episodes. The overall signficance of the GRB detection is
$\sim24\,\sigma$.

The two main pulses have different temporal profiles, including different burst
asymmetry in terms of the ratio of the rise time to the decay time. These
differences eliminate the scenario of a single pulse being seen twice due to
strong gravitational lensing.

The resulting background-subtracted light curve is at Figure\,\ref{fig:int_2lc}.
The duration of the burst is $T_{90} = 28.5\pm0.6$\,s, and the peak flux in the
1\,s window at $t_0+19$\,s was 351.72\,photons\,s$^{-1}$. We carefully
inspected the ISGRI signal during the following three minutes after trigger and
found a hint of emission by $T_\mathrm{0}+\sim 120$\,s. 

Among the considered spectral models (power law, the Band model, cut-off
power law (CPL), and black-body + power-law), the spectrum of the burst is best
fit with the CPL model with an extra component necessary to explain a high energy excess at about 220\,keV. However, the spectral properties of the high-energy component are difficult to constrain due to the low number of counts
above 100~keV. The model selection was based on the Akaike information
criterion \citep{Akaike74}, $AIC = 2d + C$, where $d$ is the number of degrees
of freedom of the model and $C$ the {\it cstat} Poisson log-likelihood obtained
from the fitting routine in {\it xspec}. The photon index of the CPL model is
$\alpha = -0.7\pm0.3$, the peak energy of the spectrum $E_p = (2-\alpha)E_0$
in the observer frame is $54\pm 9$\,keV, and the total fluence was $1.75\times
10^{-8}$\,erg\,m$^{-2}$\,s$^{-1}$ in the energy range of $20-200$\,keV. The
known redshift therefore implies a total isotropic energy of $E_\mathrm{iso} =
3.6\times 10^{51}$\,erg. We detected no spectral lag in either of the two
bright pulses with an estimated upper limit of $\tau<150$\,ms.

To summarise, INTEGRAL saw GRB\,190919B as a relatively faint and soft long GRB
typical in most of its aspects. An exception to this is a possible detection of a high energy spectral component at $\sim$200\,keV.

\begin{figure}
    \centering
    \includegraphics[width=\hsize]{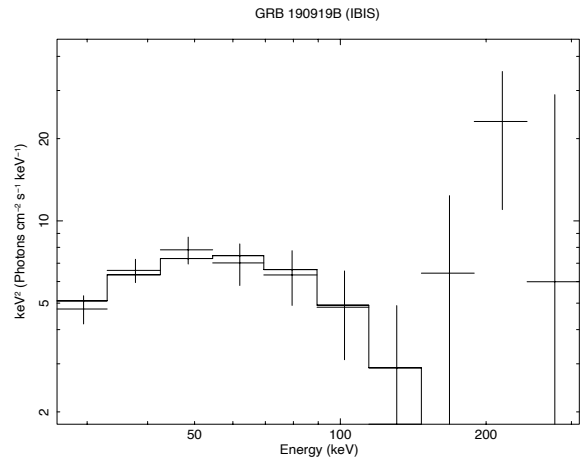}
    \caption{INTEGRAL/ISGRI (20~keV -- 250~keV) spectrum fit with a CPL model.\label{fig:int_sp}}
\end{figure}

\subsection{Ground-based optical}

\begin{table*}[t!]
\begin{center}
\begin{tabular}{ccccccl}
\hline

 $t-t_0$  [s]& exp [s]& filter& M&  $\Delta$M &  $E_\mathrm{flt}$$^\mathrm{a}$ & source\\ \hline
 407 & 388    & J  & 16.33  & 0.05   & 0.030&  \citet{GCN25789}\\ 
 407 & 388    & H  & 16.06  & 0.05   & 0.019&  \citet{GCN25789}\\ 
 407 & 388    & Ks & 15.92  & 0.07   & 0.012&  \citet{GCN25789}\\ 
 15594   &  30    & r  & 20.10  & 0.02   & 0.087& \citet{GCN25792}\\ 
 67932   & $28 \times 30$ & i' & 20.60  & 0.14& 0.066& \citet{GCN25796}\\ 
 67932   & $28 \times 30$ & r' & 21.48  & 0.18& 0.087& \citet{GCN25796}\\ 

\hline
\multicolumn{7}{l}{ \textit{Note:}
$^a$ Galactic extinction in the given filter according to \citet{Schlegel1998}}

\end{tabular}
\end{center}
\caption{Photometric measurements of the optical afterglow collected from GCN circulars \label{GCNMag}}
\end{table*}

\begin{figure}[b!]
\centering
\resizebox{!}{\hsize}{\includegraphics{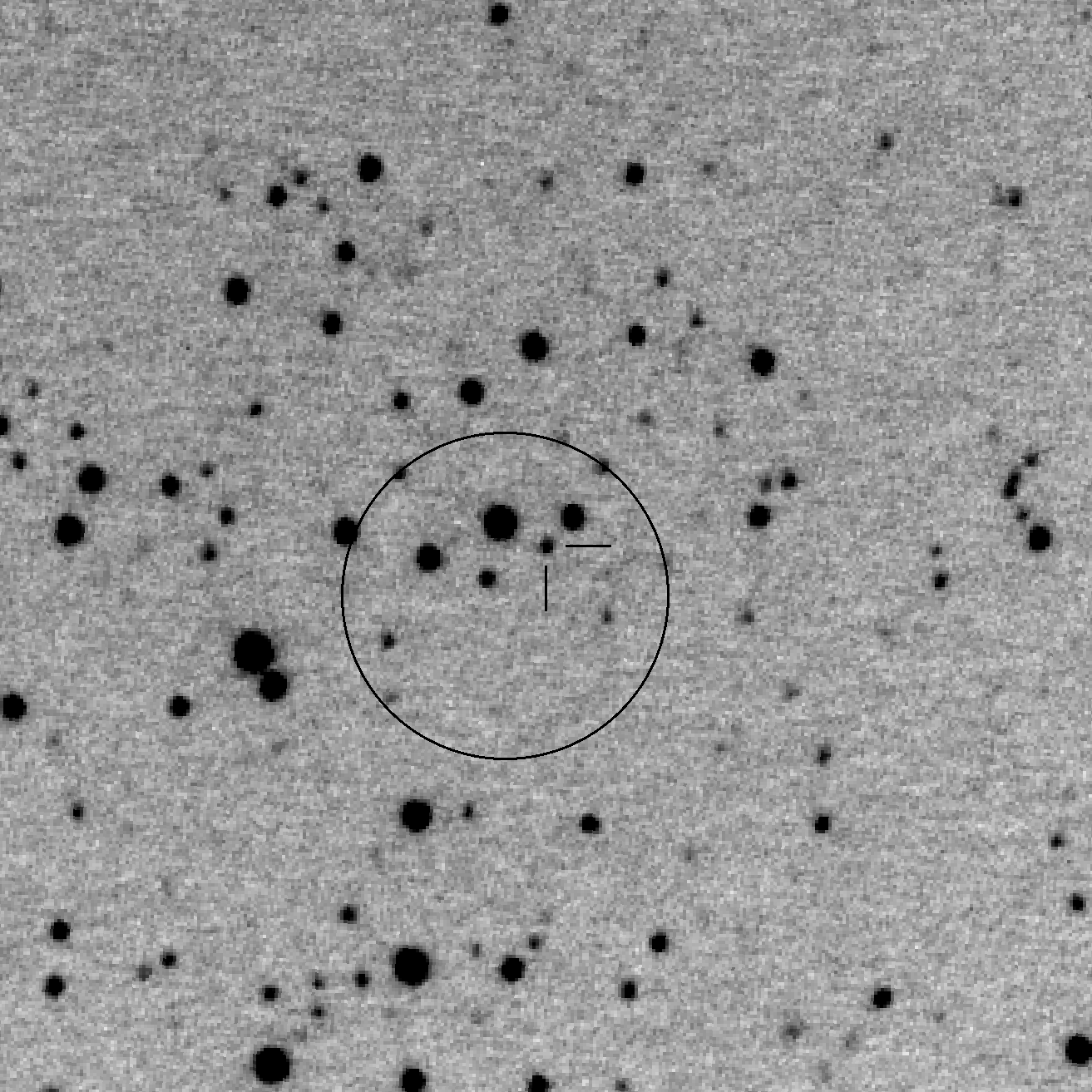}}
\caption{Image of the optical afterglow of GRB 190919B obtained by the
telescope FRAM, this image shows $10'\times10'$ with north up and east to the
left. The 1.5' {\it INTEGRAL}/IBAS error box is marked with a circle.
\label{fig:snimek}}
\end{figure}

The robotic FRAM (Ph/Fotometric Robotic Atmospheric Monitor) telescope is run
by the Institute of Physics of the Czech Academy of Sciences in Prague and is
primarily used to monitor atmospheric transparency at the Pierre Auger
Observatory in Argentina \citep{AugerFRAMnew}. In the case of reception of a
gamma-ray-burst satellite alert \citep{Bacodine}, the system is automatically
repointed towards it and obtains a pre-defined set of exposures of the burst sky
location. Since its installation in 2005, the telescope has observed tens of
such alerts, with a few notable optical afterglow detections, such as the 10.2
mag afterglow of GRB\,060117, one of the brightest afterglows ever detected
\citep{Mates060117}. The hardware configuration has changed several times, but
at the moment of GRB\,190919B it consisted of a 0.3\,m telescope with a
60'$\times$60' field of view and a 7$^\circ\times$7$^\circ$ wide-field camera.

On September 19, 2019 at 23:46:58.0 UT --- 34.6\,s after the GRB trigger ---
the {\it INTEGRAL}/IBAS \citep{GCN25788} alert n.~8377.0 was received, the
telescope interrupted observations, started to slew, and 53.5\,s after the burst
at 23:47:16.9 UT it started obtaining a set of 20\,s unfiltered exposures; then,
it continued with a set of 60\,s $R$-band frames. The CCD camera of the
FRAM telescope was set up so that it would read out only the central part of the chip
with binning $2\times2$. The images have resolution of $1024\times1024$\,pixels
and a $30'\times30'$ field of view. The observations were promptly reported
\citep{GCN25794}. 

Furthermore, 3.33\,h after the trigger, the 60cm BOOTES-5/JGT (Burst Observer and Optical Transient Exploration System, Javier Gorosabel Telescope) robotic telescope at
Observatorio Astron\'{o}mico Nacional of San Pedro Martir automatically
responded and obtained 64$\times$60\,s unfiltered exposures.
Later, we observed the same afterglow with the BOOTES-3/Yock-Allen telescope at
New Zealand \citep{bootes2012}. The observation started $\sim$ 9\,h after the
gamma-ray burst and obtained two sets of 60\,s unfiltered exposures. The
preliminary analysis of the two BOOTES observations was published by
\citet{GCN25798}.
Full observation logs are listed in Table\,\ref{tab_framdata}.
    
\begin{figure}
\centering
\includegraphics[width=\columnwidth]{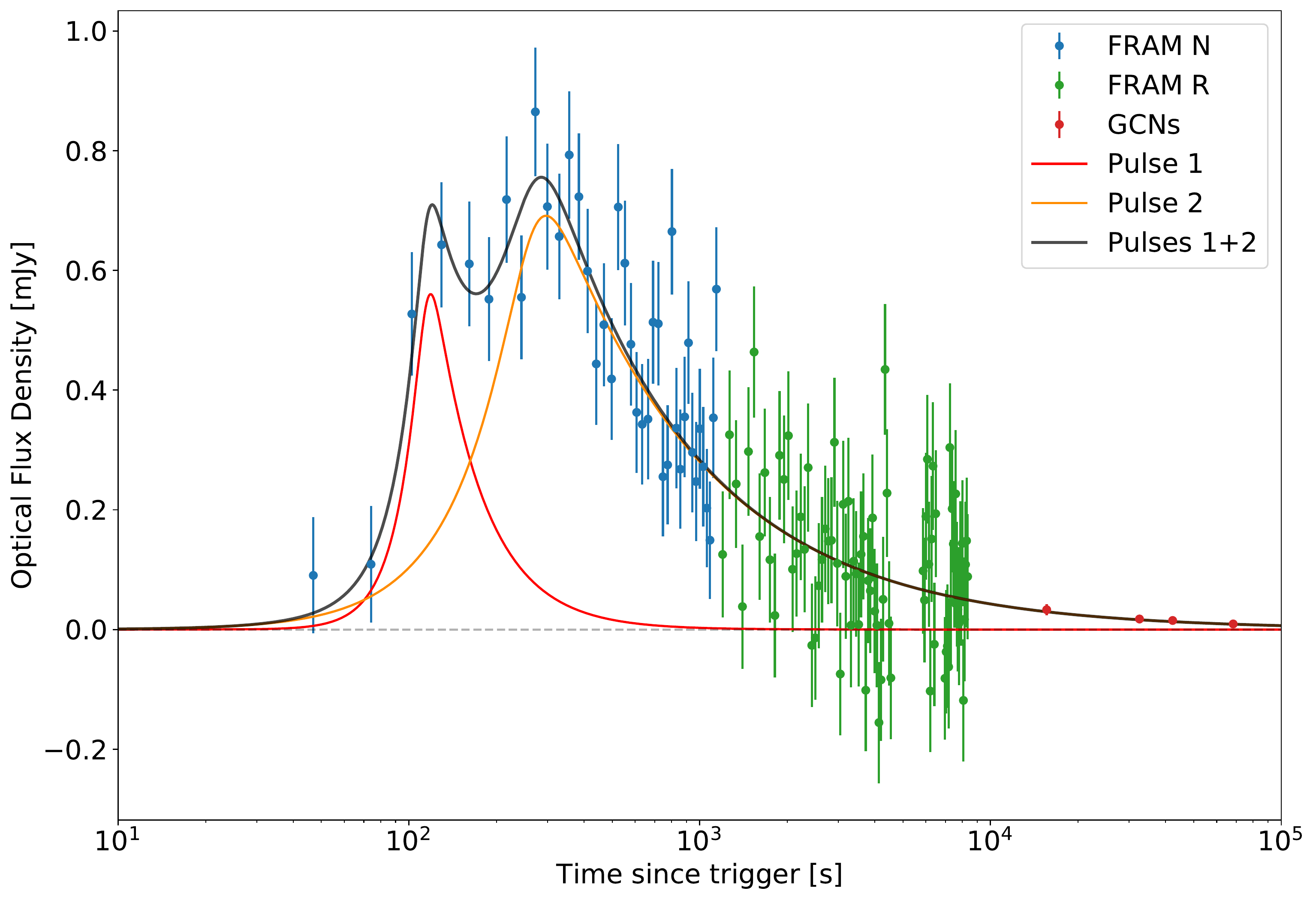}
\caption{Optical light curve of GRB 190919B, fitted with an empirical model
consisting of  a superposition of two smoothly broken power-law functions (see
Section\,\ref{sec:lc_fitting}) to describe two pulses as discussed in
Section~\ref{sec:discussion}.  \label{fig:lc}}
\end{figure}

\begin{figure}[b!]
\centering
\includegraphics{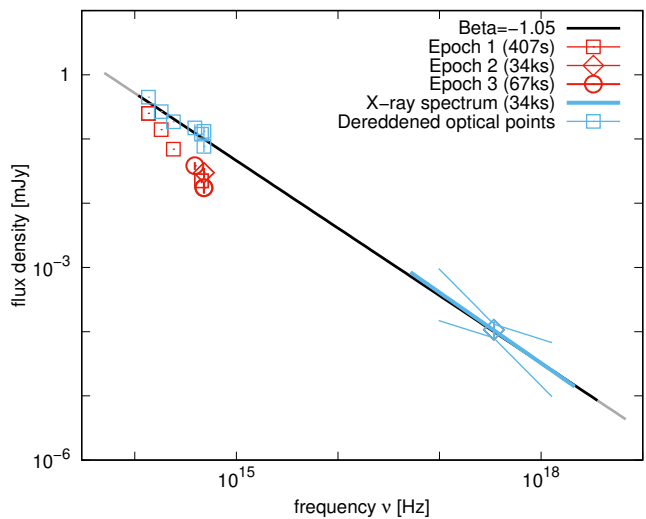}
\caption{Spectral energy distribution (SED) of GRB\,190919B afterglow. Three
different epochs were collapsed to provide the most comprehensive image
available. The optical points were scaled so that they would represent the
afterglow at 34\,ks after the trigger. All observations are in agreement with a
spectral slope of $\beta=1.05$ derived from a fitted late-decay rate. X-ray point
to the right is displayed together with its respective spectral slope and
including the slope uncertainty (hence the butterfly). Frequencies are in the
observer frame, the optical flux of the afterglow was corrected for Galactic
extinction, as was the X-ray spectrum for interstellar hydrogen absorption. The
blue optical and infreared 
points were also corrected for host galaxy dust absorption of
$E(B-V)=0.28$. \label{fig:sed}} 
\end{figure}

\section{Analysis}

In the earliest images (except the first two), the afterglow was detectable in
single exposures, and after $\sim$ 20 minutes it disappeared; however, by co-adding the
exposures, it was possible to detect it until 120\,min post trigger. By
co-adding images from the telescope BOOTES-3, we were able to detect the optical
source as late as 9.54\,h after the GRB, allowing us to better estimate the
temporal decay rate. 

\begin{figure}
\centering
\includegraphics{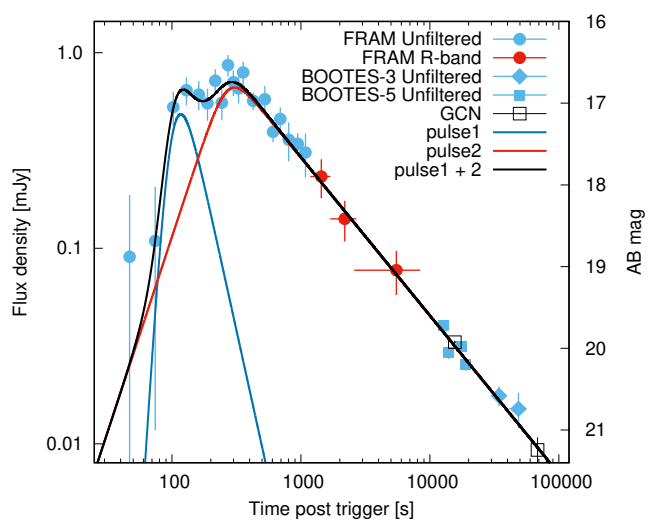}
\caption{Optical light curve of GRB 190919B in log-log plot in which the used
power laws show as straight lines. The fitted, smoothly broken power-law pulse
functions (blue, red) take the form of a hyperbola. The final model is a
superposition of these two pulses (black line). The points shown are the
photometric points from Table\,\ref{tab_framdata}. We note that the fitting was
performed in linear space and with single-image fluxes, as shown in
Figure\,\ref{fig:lc}.\label{fig:fitcolor}}
\end{figure}

An identification chart with the optical afterglow of GRB\,190919B marked is
shown in Figure\,\ref{fig:snimek}, and the obtained photometric points are listed
in Table\,\ref{tab_framdata}. However, in order to properly treat low fluxes
and errors, we did not use the FRAM points for fitting; instead, we blindly
measured signal within the aperture in every single frame and performed the
fitting with many imprecise points rather than few precise ones. This means that we did
not follow the common {\it \emph{add - detect - centre - measure}} procedure used for
CCD photometry of faint astronomical objects. This procedure also permits us to
use a robust fitting algorithm to identify and ignore frames that might
otherwise influence the co-added photometric signal. The light-curve fitting was
performed in linear space (i.e. not in the logarithmic magnitude space) with
these points (see Figure\,\ref{fig:lc}). The light curve from both telescopes
complemented with the photometric points collected from GCN circulars is shown
in Figure\,\ref{fig:fitcolor}. 

The earliest images from FRAM were obtained before the afterglow reached its
maximum brightness. The maximum can be seen by about 120 -- 420\,s and reaches
$\sim$ 16.5\,mag. From 600\,s on, only a simple power-law decay is observed. 

To quantify the possible host-galaxy contamination of the observed photometric
data, we searched for a host in the archive images of the Legacy survey
\citep{legacy}. There seems to be no detection with an estimated limit in the
range of $r' \sim 25$. The faintest known GRB\,190919B afterglow detection
with $r' = 21.48$ by \cite{GCN25796} should therefore have at most 4\% light
from the host. 
   
\subsection{Light-curve fitting}
\label{sec:lc_fitting}

\begin{figure} \centering \includegraphics[width=\columnwidth]{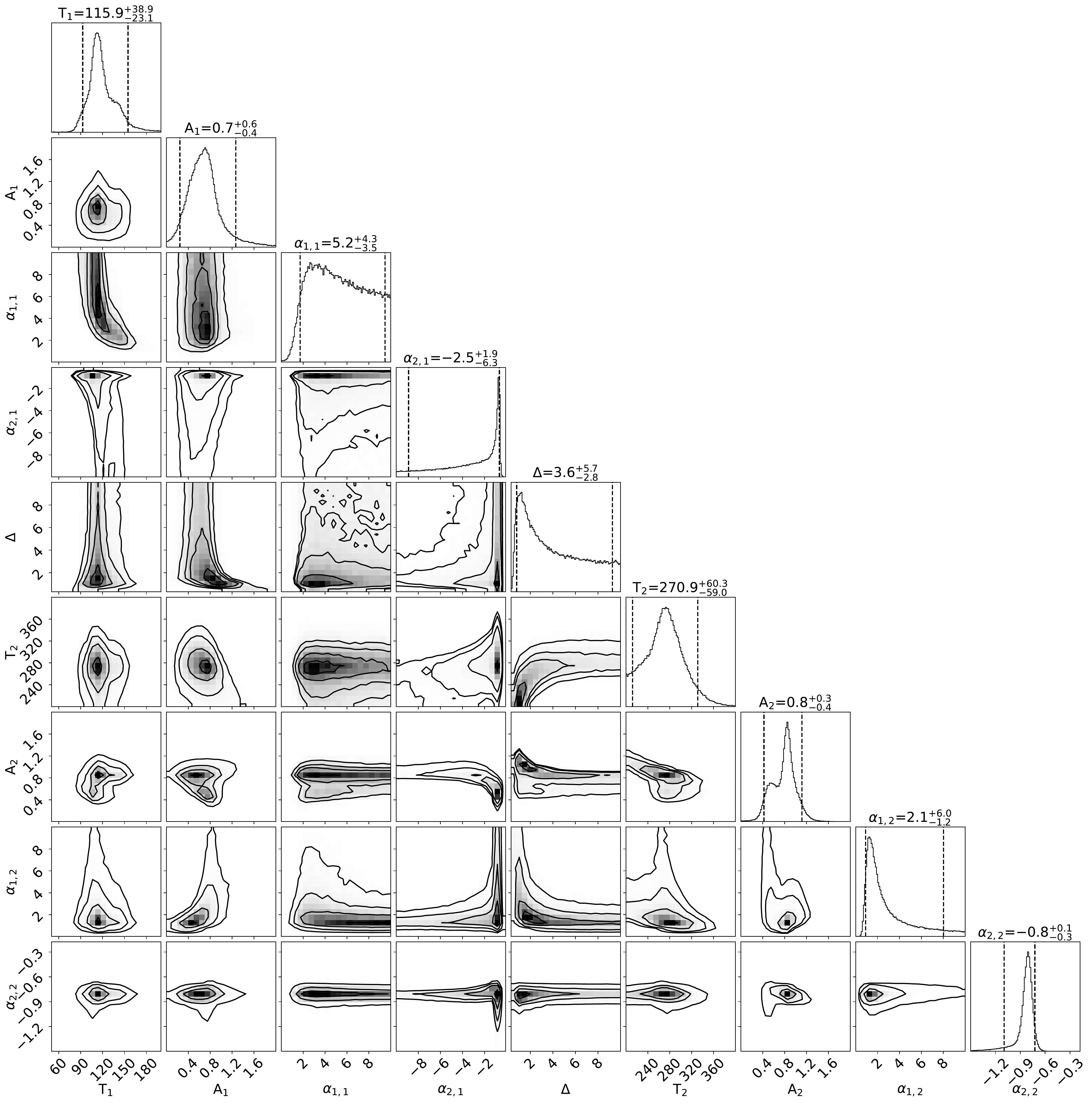}
\caption{ Parameters for the empirical model fitting of the light curve with
two smoothly broken power laws as described in Section~\ref{sec:lc_fitting}.
\label{fig:corner}} \end{figure}

In order to characterise the complex shape of the optical light curve shown in
Figure~\ref{fig:lc}, we fitted it with an empirical model consisting of two
smoothly broken power-law functions of the following form:
\begin{equation} f(t) = A \cdot \left(
\left(\frac{T_{0}}{t}\right)^{\alpha_1\Delta} +
\left(\frac{T_{0}}{t}\right)^{\alpha_2\Delta} \right)^{-\frac1{\Delta}}
,\end{equation}
which peaks at $t=T_\mathrm{0}$ and shows the asymptotic behaviour of power laws with
slopes of $\alpha_1$ and $\alpha_2$ to the left and right of the peak,
respectively. The smoothing parameter $\Delta$ defines how sharp the
transition is between the two power-law segments. In our analysis, we assumed, for
simplicity, that this smoothing parameter is the same for both peaks. We also
assumed uniform priors with reasonable limits (see
Table~\ref{tab:bayesian_fit}) for all the parameters of the resulting model.
Finally, we used Goodman $\rm \&$ Weare’s affine invariant Markov chain
Monte Carlo (MCMC) ensemble sampler \citep{Goodman2010} implemented by the
Python package {\sc emcee} \citep{Foreman-Mackey2013} to explore the parameter
space. To ensure fit convergence, we allowed the MCMC to run until the number
of steps exceeded one hundred times the maximum of the auto-correlation length
of all parameters. Then, we used 5\%, 50\%, and 95\% quantiles of marginalised
parameter distributions in the samples to derive their best fit values and 90\%
confidence limits.  The resulting regions of acceptable parameters are shown in
Figure~\ref{fig:corner} and summarised in Table~\ref{tab:bayesian_fit}.
Figure~\ref{fig:lc} shows the model corresponding to best fit parameters
overplotted on the data used for the fit. Reduced $\chi^2$ of this model is
1.29 (for 116 degrees of freedom), thus ensuring that the model is indeed
adequate to the data, and the confidence limits mentioned above are reliable.

\subsection{Spectral energy distribution}

\label{sec:sed}

Using the fitted temporal behaviour, we combined the available data from 407\,s
\citep[infrared (IR) observations,][]{GCN25789}, 34\,ks \citep[Swift
X-ray,][]{Evans2007,Evans2009,Evans2010} , and 67\,ks \citep[late
$r'$+\,$i'$,][]{GCN25796} to create a 
SED
of the afterglow emission.

The X-ray data we used (see Table\,\ref{tab:xrt}) were corrected for
interstellar hydrogen absorption using the hydrogen absorption column 
$N_\mathrm{H}=2.75\times10^{20}$ cm$^{-2}$.
The optical data were also corrected for the Galactic dust absorption
according to the Galactic 
reddening maps of \citet{Schlegel1998}. Both X-ray
and optical data were converted to the same units of measurement (janskys) for the
purpose of comparison.

Despite the relative distance between the three epochs, a common, simple
power-law spectrum can be constructed. We fitted the optical and IR points for dust
extinction in the host galaxy, which we approximated using the values adopted
from \citet{Cardelli1989}. Under the assumption that the broadband spectrum is a
simple power law, we obtained a rough estimate of the dust reddening of
$E(B-V)=0.28 \pm 0.05$, together with a spectral slope of $\beta_\mathrm{fit} =
1.05 \pm 0.07$.  This result is in agreement with both the X-ray spectrum and
the expected spectral slope of $\beta = 2\alpha/3 + 1/2 = 1.05$, as derived
from the fitted late temporal decay rate $\alpha_\mathrm{2,2}$ (see
Figure\,\ref{fig:sed}). 

%


\begin{table}
\begin{center}
\begin{tabular}{ c c c c }
\hline
$t-t_0$  [s]&exp [s]&filter&Flux [mag]\\ \hline
& & &\\
\multicolumn{4}{c}{FRAM/Auger} \\
\hline
46.9 & 20 & N & 19.00 $\pm$ 0.90 \\
74.1 & 20 & N & 18.80 $\pm$ 0.70 \\
102.4 & 20 & N & 17.09 $\pm$ 0.15 \\
129.6 & 20 & N & 16.87 $\pm$ 0.12 \\
161.3 & 20 & N & 16.93 $\pm$ 0.14 \\
188.5 & 20 & N & 17.04 $\pm$ 0.15 \\
216.6 & 20 & N & 16.75 $\pm$ 0.14 \\
243.8 & 20 & N & 17.03 $\pm$ 0.15 \\
272.2 & 20 & N & 16.55 $\pm$ 0.10 \\
299.4 & 20 & N & 16.77 $\pm$ 0.12 \\
328.8 & 20 & N & 16.85 $\pm$ 0.11 \\
356.0 & 20 & N & 16.64 $\pm$ 0.11 \\
426.2 & 20 & N & 17.02 $\pm$ 0.08 \\
525.1 & 4 $\times$ 20 & N & 16.99 $\pm$ 0.18 \\
607.3 & 3 $\times$ 20 & N & 17.41 $\pm$ 0.09 \\
692.0 & 3 $\times$ 20 & N & 17.25 $\pm$ 0.14 \\
803.3 & 3 $\times$ 20 & N & 17.77 $\pm$ 0.09 \\
951.4 & 4 $\times$ 20 & N & 17.68 $\pm$ 0.08 \\
1071.4 & 9 $\times$ 20 & N & 17.93 $\pm$ 0.20 \\
1422.6 & 7 $\times$ 60 & R & 18.13 $\pm$ 0.21 \\
2189.8 & 14 $\times$ 60 & R & 18.52 $\pm$ 0.22 \\
5558.4 & 57 $\times$ 60 & R & 19.34 $\pm$ 0.21 \\
& & &\\
\multicolumn{4}{c}{BOOTES-5 (Javier Gorosabel telescope)} \\
\hline
 12697.2  & 16$\times$60    & N  & 19.89  $\pm$ 0.08 \\
 14019.4  & 16$\times$60    & N  & 20.24  $\pm$ 0.08 \\
 17739.2  & 15$\times$60    & N  & 20.16  $\pm$ 0.06 \\
 19064.2  & 16$\times$60    & N  & 20.38  $\pm$ 0.08 \\
& & &\\
\multicolumn{4}{c}{BOOTES-3 (Yock-Allen telescope)} \\
\hline
31859 & 18$\times$60 & N & 20.79 $\pm$ 0.12 \\
46359 & 12$\times$60 & N & 20.95 $\pm$ 0.23 \\
\hline
\end{tabular}
\end{center}
\caption{Photometric measurements of the optical afterglow of
GRB~190919B}\label{tab_framdata}
\end{table}

\begin{table}[]
    \begin{center}
    \begin{tabular}{cccl}
    \hline
     & Best & 90\% CL &  Prior \\
    \hline
    
    T$_1$ & 116 & 93 ... 155 & uniform [50, 200] \\
    A$_1$ & 0.67 & 0.25 ... 1.27 & uniform [0, 2] \\
    $\alpha_{1,1}$ & 5.2 & 1.7 ... 9.5 & uniform [0, 10] \\
    $\alpha_{2,1}$ & -2.5 & -8.9 ... -0.7 & uniform [-10, 0] \\
    $\Delta$ & 3.6 & 0.8 ... 9.3 & uniform [0, 10] \\
    T$_2$ & 271 & 212 ... 331 & uniform [200, 400] \\
    A$_2$ & 0.80 & 0.4 ... 1.1 & uniform [0, 10] \\
    $\alpha_{1,2}$ & 2.1 & 0.9 ... 8.0 & uniform [0, 10] \\
    $\alpha_{2,2}$ & -0.8 & -1.1 ... -0.7 & uniform [-1.5, 0] \\
\hline
    \end{tabular}
    \end{center}
\caption{Parameters for the empirical model fitting the light curve with two
smoothly broken power laws, as described in Section~\ref{sec:lc_fitting}.  Best
values correspond to the medians of marginalised posterior probability.  The
last column shows the priors used to perform the Bayesian fit.
\label{tab:bayesian_fit} }
\end{table}

\subsection{Search for late, high-energy emission}

\begin{figure} \centering
\includegraphics[width=0.5\textwidth]{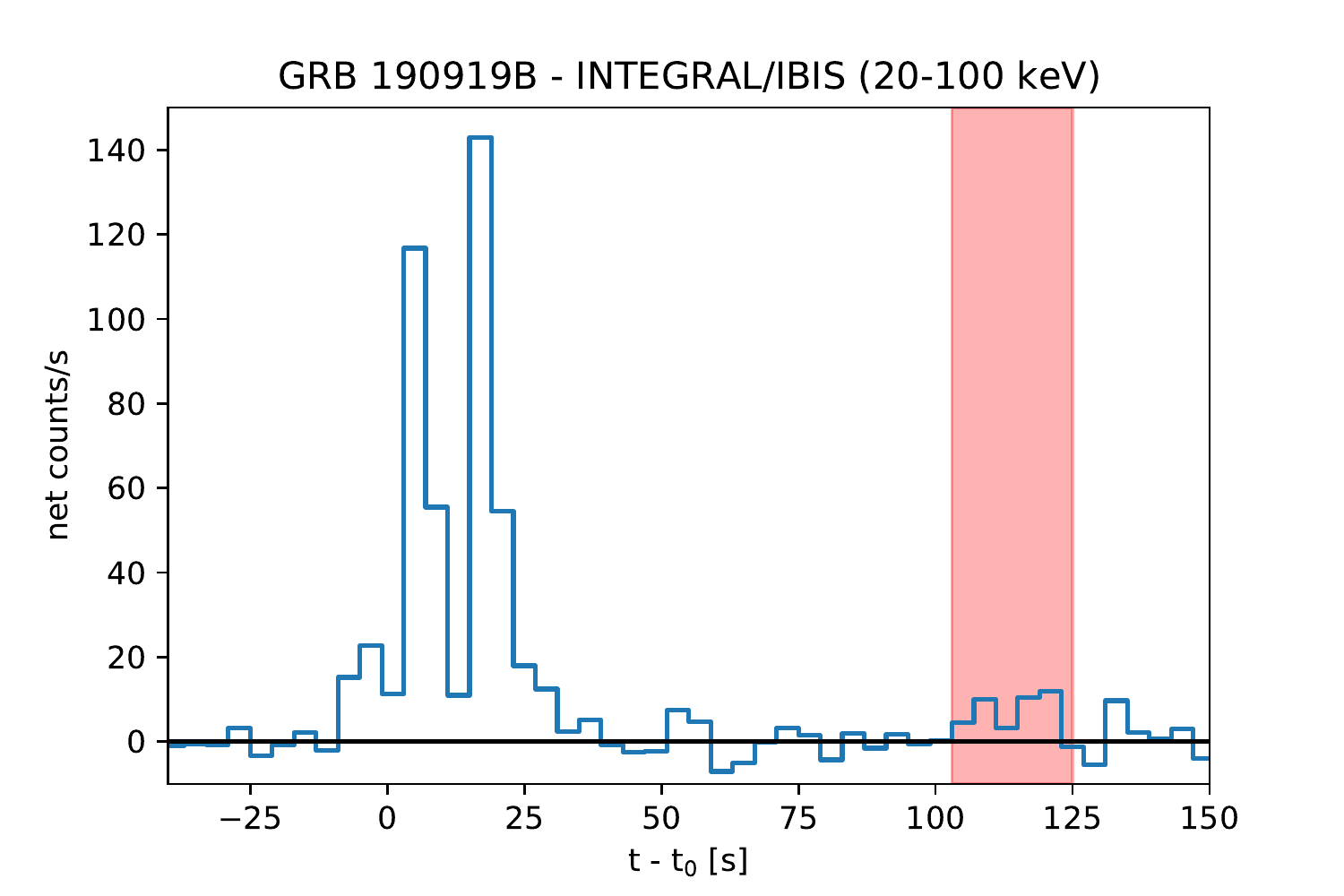} \caption{Possible,
faint, $\sim 3\sigma$ flare highlighted by the flare-finding algorithm applied
blindly on the light curve of GRB 190919B in the 20-100~keV energy range with
a coarse binning of 4~s.\label{fig:ffidner_at_120s}} \end{figure}

A careful search for the late, high-energy INTEGRAL/ISGRI emission was performed
around the time $(t-t_0) \approx 200$\,s, at which the optical afterglow peaks.
A flare-finding algorithm that works with flexible time binning (see
\citet{2021ApJ...921L...3M}) suggested a possible $3.4\,\sigma$ faint detection
in the 20--100 keV light curve between $103 \rm{~s} < (t-t_0) < 126$~s (see
Fig.~\ref{fig:ffidner_at_120s}). To further investigate the significance of the
potential signal, we inspected the distribution of the counts on the detector
as a function of pixel illumination fraction (PIF) from the source at the
position of GRB~190919B during the time interval selected by the flare-finder
routine. The hypothesis of the existence of the signal and the background were
tested against the hypothesis with only the background contribution similarly
to the method described in \citet{2021ApJ...921L...3M} (Rigoselli et al., in
prep.). A Bayesian approach using the MCMC fitting yields the significance of
the source with 3.11$\sigma$, calculated according to $ S = \sqrt{ -2 \ln
\lambda} $ , where $\lambda \equiv \mathcal{L}(b)/\mathcal{L}(s+b)$ is the
likelihood ratio of the likelihoods of the observed data given the hypothesis
with the background only, resp. with the background plus the signal as the
underlying model \citep{1983ApJ...272..317L}. Further investigation of the
ISGRI detector plane revealed the necessity to correct the counts for a bad
pixel that escaped the OSA bad pixel selection filter. This reduced the
significance of the detection to $2.96\sigma$.

For comparison, the GRB itself within the duration of the $T_{90}$ had
a significance of $23.9\sigma$, while randomly selected regions of the same duration
as the late emission candidate containing no clear signal at different times
before and after the flare showed only a $0.06 - 0.10\sigma$ detection at best using
this method.
Using a rough estimate of $\sim$25\,$\mu$Jy at $\sim$50\,keV for this emission,
we obtain a spectral slope of 0.3, which is significantly harder than that of the
later optical afterglow. 

\section{Discussion}

\label{sec:discussion}
 
\subsection{The first pulse}

Our preferred interpretation of the steep rising and the first pulse is an
ongoing activity of the internal engine. This way, we obtain a simple explanation
for the steep rise - the flare would follow its own time frame and it would not
be bound to the shell dynamic of the afterglow. 

The contemporaneous X-ray and optical peaks point towards a common origin
similar to that of an XRF 071031 \citep{Kruhler2009ApJ}; this was also a soft GRB, and
the spectral X-rays-to-optical ratio was similar ($\beta_\mathrm{X-opt} \sim
0.3$). The similarity makes us believe that what we have seen was, in fact, a
late activity of the inner engine, which, with time, slowly softens and shifts
its spectral peak towards longer wavelengths.

A competing interpretation for this pulse may be a reverse shock (RS) emission
related to the later-observed afterglow, and while it cannot be completely
disregarded, it is disfavoured by the ISGRI hard X-ray detection. Also, the
RS timescale is expected to be different -- it is expected to start
rising immediately when the ejecta hits the interstellar matter (ISM) \citep{Zhang2003} -- that is
earlier than observed. The temporal index of the rise of this pulse
($\alpha_{1,1}\simeq 5.2$ with respect to $T_0$) is also steeper than expected
for a reverse shock emission \citep{SariPiran1999}. 

On the other hand, \citet{martincarillo2014} observed a strong optical pulse at
a similar time, coincident with the end of the gamma-ray emission for a much
brighter GRB\,120711A, and interpreted it as a reverse shock emission. So, while
noting that the high-energy emission related to our pulse is much weaker, a
RS+FS scenario similar to their interpretation is also possible. While we prefer the scenario of the broadband optical+X-ray flaring activity, a
definitive decision on the nature of the pulse is impossible with the available
data.

\subsection{The second pulse and closure relations}

For the second fitted pulse, we propose an interpretation of a hydrodynamic
maximum of an expanding-shell afterglow emission. For the optical afterglows as interpreted by the relativistic fireball model,
the emission during the decay depends on time and frequency as $F\sim t^\alpha
\nu^\beta$, where the indexes $\beta$ and $\alpha$ are bound together with an
electron distribution parameter $p$ (which itself depends on microphysical
parameters). The precise relations for the $\alpha$ and $\beta$ parameters
depend, however, on the regime of the shockwave and the medium-density profile
it is spreading into.

The rising temporal index of this pulse $\alpha_{1,2} \simeq 2.1$ is a value
expected for a rising edge of an optical afterglow. Our fitted final optical decay is $\alpha_2=0.81\pm0.10$. If this decay
represents slow-cooling ejecta expanding into constant density profile,
following \citet{Sari1998} we could expect this decay to be related with the
value of the electron-distribution index 
$p$ as $\alpha=3(1-p)/4$. with $p = 2.08\pm0.13$.
The {\it Swift-XRT} photon index $\beta_\mathrm{X}=1.1$ agrees well with the
model relation $\beta=p/2$. Furthermore, with a host extinction
$E(B-V)_\mathrm{host}=0.28$ (see Sect.\,\ref{sec:sed}), the X-ray-to-optical
spectral slope becomes perfectly compatible with the spectral slope
$\beta=1.05$ derived from the afterglow theory. %

As noted above, the late Swift-XRT observation obtained 128\,ks post-burst is
considered unreliable as it is too weak. The associated values of the decay
rate of $\alpha_\mathrm{X} = 1.30{{+0.5}\above 0pt {-0.4}}$ and the
X-ray-to-optical spectral slope of $\beta_\mathrm{OX,late}=0.99$ may point towards
a change of regime, but deeper inspection of the late-time behaviour is
impossible with the available data.

Testing other possible combinations of conditions (fast cooling, wind profile,
post jet-break decay) provides unrealistically low values of $p<<2$ and no
compatibility between spectral and optical indices. 

\subsection{Initial Gamma factor}

For the fitted values, we can estimate the initial Lorentz factor of the ejecta
$\Gamma_0$ (similarly to~\citealt{Molinari2007}), for which we have:
\begin{equation} \Gamma_0= 2 \Gamma(t_\mathrm{peak}), \end{equation}
where
\begin{equation} \Gamma(t_\mathrm{peak}) \approx 160 \left[ \frac{E_{\gamma ,
53}(1+z)^3}{\eta_{0, 2} n_0 t_\mathrm{peak,2}^3} \right]^{1/8}, \end{equation}
$E_{\gamma }=E_{\gamma , 53}10^{53}$~erg is the overall isotropic energy, and
$t_\mathrm{peak, 2}=t_\mathrm{peak}/100$~s is a corrected time of the maximum.
%
The values $\eta_{0, 2}$ and $n_0$ are unknown but are assumed to be close to
unity and the resulting Lorentz factor depends only weakly on them. Using the fitted value of$T_\mathrm{2}=271.1^{+33.5}_{-36.3}$\,s and the
isotropic energy of $E_\mathrm{iso} = 3.6\times 10^{51}$\,erg, we obtain the
initial Lorentz factor value of $\Gamma_0 = 250$,
in agreement with values expected for the relativistic fireball model
($\Gamma_0$ in range of 50 to 1000, see \citealt{Piran2000})

\section{Conclusions} 

GRB\,190919B was a 26\,s gamma-ray burst detected at a redshift of
$z=3.225$ by INTEGRAL. It consisted of one faint and two bright gamma-ray
pulses, followed by a marginally significant hint of a flaring event
$\sim$120\,s post-trigger. The spectrum of the burst was 
relatively soft, but shows an apparent excess of gamma-ray emission 
at~200\,keV during the brightest pulse.

Following its localisation, we obtained early photometry of the optical
afterglow with three robotic telescopes: FRAM, BOOTES-3, and BOOTES-5. The data
we collected permitted us to construct a light curve of the afterglow, and,
complemented with publicly available information, we were also able to construct its
spectral energy distribution.
The optical light curve rose steeply $\sim
100\,s$ after the trigger (almost 10$\times$ in brightness), and eventually it
reaches a maximum of 16.5\,mag. The light curve then started to become fainter and
settles at a power-law decay rate of $\alpha_2=0.81\pm0.10$ until it faded
away.

We interpret the steeply rising afterglow light curve as the superposition of
two pulses of different physical natures. The first pulse can be plausibly
interpreted as a flare corresponding to internal engine activity. This scenario
is supported by a hint of hard X-ray emission detected simultaneously to the
fitted pulse in the INTEGRAL/ISGRI data. 

The second pulse is interpreted similarly to other GRBs as an onset of the
afterglow emission and a forward shock (FS) of an ejecta colliding with a constant
density interstellar matter. The initial gamma factor corresponding to the
delay between the GRB trigger and the peak of the emission is $\Gamma_0 \approx
250$. The late afterglow decay and X-ray-to-optical spectrum is consistent with
the prediction of the relativistic fireball model with an expansion in the slow-cooling regime into a constant density interstellar medium with an electron
distribution parameter of $p=2.08\pm0.13$. 

\begin{acknowledgements}
We would like to thank the Pierre Auger Collaboration for the use of its
facilities.
Further, we would like to thank the Lauder Atmospheric Research Station and dr.
Richard Querel for hosting and support of BOOTES-3.
The operation of the robotic telescope FRAM is supported by the grant of the
Ministry of Education of the Czech Republic LM2018102. 
The data calibration and analysis related to the FRAM telescope is supported by
the Ministry of Education of the Czech Republic MSMT-CR LTT18004, MSMT/EU funds
CZ.02.1.01/0.0/0.0/16\_013/0001402 and CZ.02.1.01/0.0/0.0/18\_046/0016010.
SK and MP acknowledge support from the European Structural and Investment Fund
and the Czech Ministry of Education, Youth and Sports (Project CoGraDS --
CZ.02.1.01/0.0/0.0/15\_003/0000437).
MT and MR acknowledge support from the Ministry of Education and Research of
Italy via project PRIN-MIUR 2017 UnIAM (Unifying Isolated and Accreting
Magnetars).
YDH acknowledges support under the additional funding from the
RYC2019-026465-I. 
AJCT acknowledges support from the Spanish Ministry Project
PID2020-118491GB-I00 and the "Center of Excellence Severo Ochoa" award for the
Instituto de Astrofísica de Andalucía (SEV-2017-0709) as well as technical
support from both NIWA Lauder and San Pedro M\'artir Astronomical Observatory
staff.
We would also like to thank an anonymous referee for helpful comments that
improved the quality of the paper.

\end{acknowledgements}

\bibliographystyle{aa}
\bibliography{aa}

\end{document}